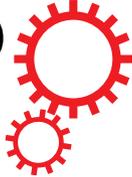

# SCIENTIFIC REPORTS

**OPEN**

# Computer keyboard interaction as an indicator of early Parkinson's disease

L. Giancardo[1,*], A. Sánchez-Ferro[1,2,3,4,5,*], T. Arroyo-Gallego[1,6], I. Butterworth[1], C. S. Mendoza[1], P. Montero[7], M. Matarazzo[2,3,4,5], J. A. Obeso[2,3,4], M. L. Gray[1,8] & R. San José Estépar[9]

Parkinson's disease (PD) is a slowly progressing neurodegenerative disease with early manifestation of motor signs. Objective measurements of motor signs are of vital importance for diagnosing, monitoring and developing disease modifying therapies, particularly for the early stages of the disease when putative neuroprotective treatments could stop neurodegeneration. Current medical practice has limited tools to routinely monitor PD motor signs with enough frequency and without undue burden for patients and the healthcare system. In this paper, we present data indicating that the routine interaction with computer keyboards can be used to detect motor signs in the early stages of PD. We explore a solution that measures the key hold times (the time required to press and release a key) during the normal use of a computer without any change in hardware and converts it to a PD motor index. This is achieved by the automatic discovery of patterns in the time series of key hold times using an ensemble regression algorithm. This new approach discriminated early PD groups from controls with an AUC = 0.81 (n = 42/43; mean age = 59.0/60.1; women = 43%/60%;PD/controls). The performance was comparable or better than two other quantitative motor performance tests used clinically: alternating finger tapping (AUC = 0.75) and single key tapping (AUC = 0.61).

Parkinson's disease (PD) is the second most prevalent neurodegenerative disorder in the western world[1]. Subtle motor manifestation can precede the clinical diagnosis by several years and continue throughout the course of the disease, however they often go unnoticed particularly in the early stages[2–4]. After this point of diagnosis, patients typically follows a progressive course leading to severe disability and shortening life span[5,6]. A number of drugs are available for symptomatic relief, including levodopa, dopamine agonists and MAO-B inhibitors[7]. These types of treatments administered by a specialist significantly lowered the risk of hip fractures, admissions to skilled nursing facility and increased survival rates[8].

An accessible way to precisely quantify PD motor signs in the patient's home has the potential to bring significant benefits to therapy management, better diagnosis and possibly earlier detection of the symptoms and enabling the development of new therapies[9,10]. The current standard to evaluate motor signs is the Unified Parkinson's Disease Rating Scale part III (UPDRS-III)[11], a compound clinical score that ascertain various motor aspects of the disease, such as rigidity, resting tremors, speech and facial expression among others. This scale requires trained medical personnel and attendance of the patient in the clinic, limiting the ease and frequency with which it can be administered. Longitudinal clinical studies measuring motor signs typically have a time resolution at least 3 months[12]. Outside the clinical study settings, patients report visits with their neurologist every 2 to 6 months[13]. Thus, the time frame over which a clinician can act on information is intrinsically many months, while many aspects of PD can fluctuate broadly in time, from hours to months, thus making higher frequency PD signs quantification tools an unmet medical need for PD management[14].

[1]Madrid-MIT M+Visión Consortium, Research Laboratory of Electronics, Massachusetts Institute of Technology, Cambridge, MA, USA. [2]HM Hospitales - Centro Integral en Neurociencias HM CINAC, Móstoles, Madrid, Spain. [3]CEU San Pablo University, Campus de Moncloa, Calle Julián Romea, 18, 28003 Madrid, Spain. [4]Centro de Investigaci´on Biom´edica en Red, Enfermedades Neurodegenerativas (CIBERNED), Madrid, Spain. [5]Instituto de Investigación Hospital 12 de Octubre (i+12), Madrid, Spain. [6]Universidad Politécnica de Madrid, Spain. [7]Movement disorders unit, Hospital Clinico San Carlos, Madrid, Spain. [8]The Institute of Medical Engineering and Science, Massachusetts Institute of Technology, Cambridge, MA, USA. [9]Brigham and Women's Hospital, Harvard Medical School, Boston, MA, USA. *These authors contributed equally to this work. Correspondence and requests for materials should be addressed to L.G. (email: gianca@mit.edu)

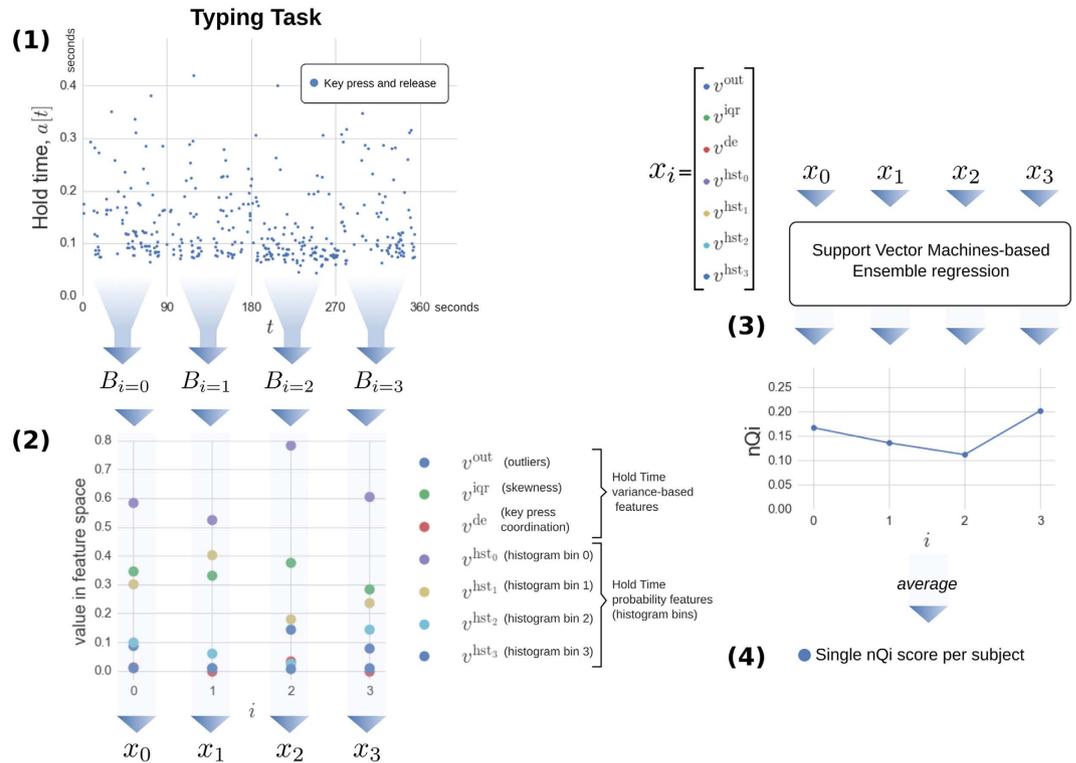

**Figure 1. Pipeline of the algorithm to generate the neuroQWERTY score (nQi) from the hold time (HT) series.** (1) The HT time events are split by non overlapping 90 seconds windows to create the $B_i$ sets. (2) From each independent $B_i$ set, a 7-element feature vector, $x_i$ is computed: 3 features that represent HT variance, and 4 features that represent a histogram of HT values. Any $B_i$ sets with fewer than 30 HT values were ignored. (3) For each feature vector, $x_i$, a single numerical score, nQi, is generated using an ensemble regression approach. Each unit in the ensemble regression includes a linear Support Vector Regression step trained on the Unified Parkinson's disease rating scale part III (UPDRS-III), the clinical score for evaluating PD motor symptoms. The Support Vector parameter estimation was done using a separate data set. A cross validation strategy with two data sets (de novo PD and early PD) was employed (see Fig. 4). (4) For the analyses herein, an average nQi score was computed for each subject tested. More information can be found in the Methods.

Digital technologies for objectively quantify PD motor signs exist and new ones are being developed[15]. One of the most frequently used is finger tapping, where subjects are asked to intermittently press buttons as fast as possible for a given time[16]. More recently, wearable inertia measurement units (IMUs) have been employed to measure information about gait, posture, tremors, bradykinesia (slow movements) and dyskinesias (involuntary movements)[17,18]. Typically, multiple sensors are applied on various areas of subject's body who is then asked to perform a particular task. IMUs can also be found in modern smartphones, which has motivated attempts to combine finger tapping and IMUs in a single device, in some cases also including voice utterances tests[19].

In the last 30 years, typing cadence (also known as keystroke dynamics) has been studied by various research groups and employed commercially as a biometric, mainly as a way to replace or strengthen passwords[20,21]. Applications to the medical field are almost non-existent, one exception are Austin *et al.* who used the typing speed in login sessions to evaluate sensory-motor speed in healthy subjects[22]. In our previous work[23], we were able to detect a pattern from keystroke dynamics that could detect a state of psychomotor impairment with a cohort of 14 healthy subjects. The state of psychomotor motor impairment was induced via a sleep inertia paradigm, i.e. the abrupt awakening during sleep. The algorithm developed was able to classify the change of state from "awaken" to "sleep inertia" with an Area Under the ROC curve of 0.93/0.91 and performance significantly superior over typing speed alone.

In this work, we demonstrate the ability to distinguish PD patients at the early stage of the disease from comparable healthy controls. We monitor their natural interactions with standard keyboards, recording the hold time (HT) occurring between pressing and releasing a key while the user is typing in a standard word processor. Then we convert the series of HTs to the numerical neuroQWERTY index (nQi) employing a novel algorithm. The system automatically learns by example the PD typing patterns by comparing the PD subjects with a control group with similar typing skills and education. Our approach does not require information about the text being typed or the actual key being pressed, only the hold time for each key (typically around 100 milliseconds)[24,25]. Figure 1 shows how the neuroQWERTY index is computed. First, the HT time series generated from the typing task is split by non overlapping windows, then 7 features are computed from the HT data set in each independent window, 4 features representing estimates of the absolute HT probability, and 3 representing different aspect of the HT variance. These features are used as the input to an ensemble regression algorithm based on $\varepsilon$-Support Vector Regression and previously trained on an independent dataset. The output of the regression algorithm is a numerical score (one score for each of the non-overlapping HT data sets). Finally, for the typing tests described here, we average the sequence of numerical scores to create a quantitative numerical index for each subject which is evaluated against the diagnostic category (PD or healthy controls) and compared to two quantitative motor tasks typically used in clinical studies.

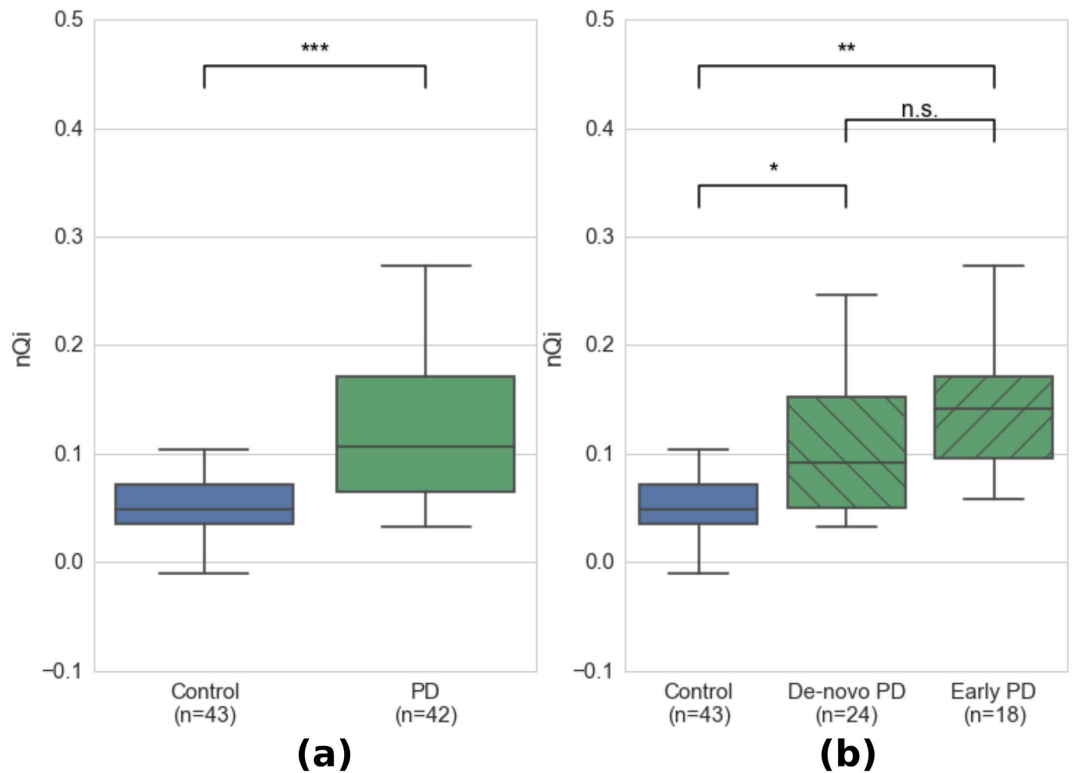

**Figure 2. Discriminative performance of nQi.** For each subject, an average nQi score was computed (as illustrated in Fig. 1) from the hold time series measured during the typing task. Box plots visualize first, third quartiles and medians; the ends of the whiskers represent the lowest (or highest) value still within 1.5-times the interquartile range. (**a**) Group level comparison between PD and controls with the combined dataset between all 43 controls and 42 PD subjects. The control group is significantly different from the PD group (p = 0.001). (**b**) Group level comparison between controls, de-novo PD subjects (recently diagnosed with PD and never taken PD medications; average time since diagnosis 1.6 years) and early PD subjects (average time since diagnosis 3.9 years; on PD medication, but no medication for the 18 hours before the typing test). Both PD sub-groups are significantly different from the controls group (de-novo/controls p = 0.022, early PD/controls p = 0.003). The statistical significance of the discriminative performance is computed with a logistic regression model including sex, age, years of education and typing skills as co-variates (see suppl. material Fig. S.3).

### Results

Figure 2(a) shows the box plots of the nQi scores computed on the typing tasks for between 43 controls and 42 PD subjects. Each box plot uses a single data point for each subject, thus allowing a meaningful comparison between the groups. After accounting for sex, age, years of education and typing speed (a common metric for typing skills) with a logistic regression model, the difference between the two groups measured with nQi are statistically significant (p-value = 0.001, see suppl. material Fig. S.3). In Fig. 2(b), the PD group is split into early PD (already-medicated PD group evaluated 18 hrs after the last dose) and de-novo (recently diagnosed group who never took any medication related to PD). These two subgroups show a statistically significant difference from the control group (de-novo/controls p = 0.022, early PD/controls p = 0.003) and similarly to the previous experiment, the logistic regression model accounted for sex, age, years of education and typing speed (see suppl. material Fig. S.3).

In Fig. 3 and Table 1, we compared the nQi scores with quantitative metrics evaluating upper limbs motor functions, i.e. finger tapping. Finger tapping is typically used in clinical trials and involves pressing one or two buttons as fast as possible for a short period of time. We evaluated two common variations of the finger tapping test: "single key tapping"[26] and "alternating finger tapping"[16]. Receivers operating characteristic (ROC) curves are used to compare the metrics. nQi showed the best classification performance with an Area under the Receiving Operating Characteristic Curve (AUC) of 0.81 (0.72–0.88 95% CI, p-value = 0.001), alternating finger tapping had a lower performance with an AUC = 0.75 (0.64–0.83 95% CI, p-value < 0.001) and single key tapping with $AUC = 0.61$ (0.51–0.71 95% CI, p-value = 0.35 ). The AUC can be interpreted as the probability that a classifier will rank a randomly chosen PD subject higher than a randomly chosen control subject. The p-values reported test the null hypothesis that the metric under scrutiny does not contribute to the separation between PD and control group in a logistic regression model accounting for sex, age and years of education (see suppl. material Figs S.3 and S.4). In order to better evaluate the discrimination performance of nQi, Table 2 shows different ROC cut-off points leading to different sensitivity/specificity ratios and accuracy. Assuming that the cost of misclassifying PD and controls is equal, nQi obtains 0.71/0.84 sensitivity/specificity and 0.78 accuracy. For completeness, Table 1 also shows the UPDRS-III scores. In our datasets, only subjects with confirmed clinical PD or lack thereof were included. Therefore, UPDRS-III, which is based on clinical evaluations, can discriminate PD subjects from controls perfectly.

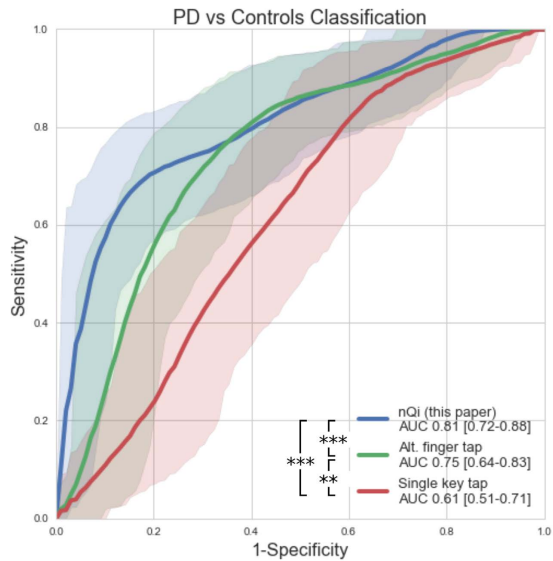

**Figure 3. Comparison of receivers operating characteristic (ROC) curves showing the classification performance of nQi (main contribution of this paper), alternating finger tapping and single key tapping on the combined dataset of 42 PD subjects and 43 controls.** The shadowed areas represent the 95% confidence intervals. In the legend, the area under the ROC curve (AUC) and the 95% confidence intervals and are shown (see Table 1 for more details). The nQi score shows the best performance in comparison with alternating finger tapping ($p < 0.001$) and single key tapping ($p < 0.001$). Alternating finger tapping and single key tapping are two quantitative measurements commonly used to evaluate motor impairment in PD studies. In our cohort, the former showed better performance than the latter ($p = 0.008$). The p-values have been computed with the DeLong's test for correlated ROC curves, which test the null hypothesis that the AUCs of two ROC curves are statistically the same.

|  | Parkinson's | Controls | Statistical Significance (unadjusted) | Statistical Significance (adjusted) |
| --- | --- | --- | --- | --- |
| $n$ (total $n = 85$) | 42 | 43 |  |  |
| Avg. UPDRS-III (std) | 20.6 (7.7) | 1.9 (1.8) | ***($p < 0.001$) | N/A |
| Avg. nQi (std) | 0.130 (0.085) | 0.060 (0.057) | ***($p < 0.001$) | ***($p = 0.001$)[a] |
| Avg. Alternating Finger Tapping (std) | 95.37 (22.01) | 128.37 (28.85) | ***($p < 0.001$) | ***($p < 0.001$)[b] |
| Avg. Single Key Tapping (std) | 162.88 (24.09) | 170.85 (16.45) | not sig. ($p = 0.08$) | *($p = 0.035$)[b] |

**Table 1. Summary of statistical tests for performance of nQi (main contribution of this paper), alternating finger tapping, single key tapping and typing speed on the combined dataset of 42 PD subjects and 43 controls.** The unadjusted statistical significance is computed with two-sided Mann-Whitney U test. [a]An additional co-variate, typing speed, was added for nQi. nQi and alternating finger tapping are the two tests that show a consistent statistical significance difference between PD subjects and controls. In the adjusted model for nQi, none of the co-variates reached statistical significance (see supplementary materials). For completeness, we also show the UPDRS-III scores. In our datasets, only subjects with confirmed clinical PD or lack thereof were included. Therefore, UPDRS-III, which is based on clinical evaluations, can discriminate PD subjects from controls perfectly. [b]The adjusted significance tests were computed with logistic regression models including sex, age and years of education as co-variates.

Table 3 shows the datasets used in the experiments. In order to limit any chance of overfitting, all nQi scores were computed with a model trained with a cross-validation strategy (see Methods). In Table 4, we show the discriminative performance in the two "folds", i.e. training nQi model on the early-PD dataset/run nQi model on De-novo dataset and vice-versa as described in Fig. 4. We obtained the following AUC: 0.92 for the Early-PD dataset, 0.77 for the De-novo dataset and 0.81 for the combined dataset. The AUC allowed us to have reliable metrics even in the presence of unbalanced number of subjects in the independent datasets.

| Misclassification cost (cost per FN/cost per FP) | Estimated cut-off point | Sensitivity | Specificity | Accuracy | TP | FN | TN | FP |
|---|---|---|---|---|---|---|---|---|
| 1/1 | 0.078 | 0.71 | 0.84 | 0.78 | 30 | 12 | 36 | 7 |
| 2/1 | 0.075 | 0.74 | 0.79 | 0.76 | 31 | 11 | 34 | 9 |
| 1/2 | 0.105 | 0.55 | 0.95 | 0.75 | 23 | 19 | 41 | 2 |

**Table 2. nQi discrimination performance with different cut-off points in the combined dataset.** TP: true positives, FN: false negatives, TN: true negatives, FP: false positives. The cut-off points have been automatically estimated by maximizing the generalized Youden Index[35] under three different misclassification costs assumptions: the cost for FN and FP is equal, the FN misclassification cost is twice the one for FP and that the FP misclassification cost is twice the one for FN.

In addition, we evaluated the discriminatory power of nQi considering each 90 seconds non-overlapping time window independently. In this case we obtained an AUC = 0.79 (0.76–0.82 95% CI) for the nQi on the combined dataset (see suppl. material Fig. S.5).

## Discussion

We are entering a new era in Parkinson's disease management. Compound clinical scores are limited by the frequency of the measurements and the subjectivity of these assessments. Hence, there is an unmet medical need for quantitative and reliable tests that can complement clinical scales for various applications (e.g. drug response evaluation, at-risk cohorts identification, among others). Our envisioned approach could complement these standards by using the unconstrained use of digital devices in a setting representative of the daily routine (i.e. an ecologically valid environment) and allow for an objective and frequent evaluation of PD motor signs. In the experiments presented, our method shows promise and can accurately discriminate an early PD population with mild parkinsonian signs from a healthy control group.

Finding consistent patterns of early Parkinson's from uncontrolled typing might appear an intractable problem, the style of typing varies greatly across subjects, an unpredictable number of pauses can be made (leading to sparse data), text typed varies greatly and so do the speed of typists. Further, the challenge was to identify a typing pattern that could be related to the pathophysiology of PD and that is not confounded by volitional action or other diseases that might impair motor function.

The problem of typing style heterogeneity was made tractable by adopting three strategies: using the Hold Time (HT) time series, automatically learning patterns from the data and considering each typing window locally. Additionally, the act of pressing and releasing a key is not influenced to a great extent by the text typed, typing speed, or typing style. We witnessed subjects who were "hunt and peck" typists having similar HTs as touch typists.

Regarding the pathophysiology of PD, typing (intended as the act of pressing and releasing keys) can be defined as a habit. Habits are greatly controlled by the basal ganglia and are more affected than goal-directed actions in PD. From the different keystroke dynamics, we selected HT: in addition to being independent of typing skills, it is largely not under conscious control. Each HT lasts on the order of 100 ms, a time so short to make implausible that subjects could consciously control it to within the 0–500 ms range where the vast majority of data lies. While a user could intentionally hold a key down for a long time, that is also not something that would happen for a prolonged period of typing. Therefore, we believe that the HT time series captures transient bradykinesia effects in typing that prevent PD subjects from lifting their fingers from keys in a consistent manner. This dynamic variance, i.e. heteroscedasticity, for motor measurements involving PD patients was already reported[27,28]. To our knowledge, our model is a first attempt to quantify this effect during a natural, uncontrolled task that did not involve visual or auditory stimuli. With nQi we strive to generate a straightforward numerical metric able to measure fine motor finger-based PD signs that could eventually be interpreted by physicians and patients alike.

We focused on early PD rather than more advanced disease stages, because it is with the early PD population where a low barrier to entry diagnostic tool could have particularly significant impact. For example, in developing treatments, particularly neuroprotective ones, clinical trial participants need to be recruited at the earliest stage possible. An easy to use diagnostic tool might also aid in lowering the number of undiagnosed PD subjects, thereby leading to adequate medical management[8]. Accordingly, we evaluated the discriminative power of our algorithm starting with a dataset that includes PD patients at the early stages when motor manifestation can be very subtle.

In our cohorts, the discriminative power of nQi scores was high even when adjusted for sex, age, education or typing skills. Still, we are far from having a fully validated diagnostic tool specific for PD motor signs. In methods like these, there is always a risk of overfitting the data such that the algorithm has limited generalizability to individuals whose PD signs are not represented in the dataset. The full external validity can be only demonstrated by a prospective study with a large sample size that encompasses the broad spectrum of motor and non-motor characteristics that are present in a sporadic PD population (such as cognition, depressive symptoms, apathy or anxiety).

Before undertaking such a study, several strategies can be used to increase confidence in the approach and mitigate the risk of overfitting. One set of strategies relates to how the training and test sets are handled. Here, we used a conservative cross-validation strategy where with two separate data sets, we train on one data set, and test with the other, and then repeat with the train and test datasets swapped. The resulting area under the ROC curve (AUC) for the combined datasets was 0.81 which is comparable or better than two other common quantitative motor performance tests to evaluate upper limbs: alternating finger tapping and single key tapping (see Fig. 3). As we and others accumulate more datasets, it will be possible to refine the nQi model with additional training data and compare it to other tests. To accelerate these lines of research we are making the datasets used in this paper available to the research community.

|  | Parkinson's | Controls | Statistical Significance |
|---|---|---|---|
| **Combined dataset** | | | |
| $n$ (total $n = 85$) | 42 | 43 | |
| Avg. Disease onset, years (std) | 2.58 (1.67) | | |
| Women # (%) | 18 (43%) | 26 (60%) | not sig. (p = 0.11) |
| Men # (%) | 24 (57%) | 17 (40%) | not sig. (p = 0.11) |
| Avg. Age (std) | 59.0 (9.8) | 60.1 (10.2) | not sig. (p = 0.53) |
| Avg. Years of Education (std) | 15.2 (4.1) | 15.3 (5.2) | not sig. (p = 0.98) |
| Avg. Typing Speed (std) | 97.91 (43.48) | 112.3 (58.7) | not sig. (p = 0.35) |
| **De-novo dataset** | | | |
| $n$ (total $n = 54$) | 24 | 30 | |
| Avg. Disease onset, years (std) | 1.60 (1.22) | | |
| Women # (%) | 10 (42%) | 16 (53%) | not sig. (p = 0.40) |
| Men # (%) | 14 (58%) | 14 (47%) | not sig. (p = 0.40) |
| Avg. Age (std) | 61.4 (10.5) | 61.8 (10.5) | not sig. (p = 0.68) |
| Avg. Years of Education (std) | 15.5 (3.8) | 14.9 (5.1) | not sig. (p = 0.55) |
| Avg. Typing Speed (std) | 97.2 (42.5) | 110.3 (59.5) | not sig. (p = 0.51) |
| **Early-PD dataset** | | | |
| $n$ (total $n = 31$) | 18 | 13 | |
| Avg. Disease onset, years (std) | 3.89 (1.23) | | |
| Women # (%) | 8 (44%) | 10 (77%) | not sig. (p = 0.08) |
| Men # (%) | 10 (56%) | 3 (23%) | not sig. (p = 0.08) |
| Avg. Age (std) | 55.9 (8.0) | 56.1 (8.6) | not sig. (p = 0.95) |
| Avg. Years of Education (std) | 14.83 (4.6) | 16.2 (5.4) | not sig. (p = 0.37) |
| Avg. Typing Speed (std) | 98.9 (45.9) | 117.0 (59.2) | not sig. (p = 0.48) |

**Table 3. The combined dataset comprises two independent data sets: De-novo dataset and Early-PD dataset.** The typing speed is computed from the dataset as the average number of keys pressed in a minute. With the exception of PD-specific scores (disease onset), the attributes of the control and PD subjects are statistically similar (using the two-sided Mann-Whitney U test), suggesting the populations are reasonably well matched. (The gender may be a confound for the Early-PD dataset by itself).

| Test dataset | Train dataset | Area Under the ROC curve (AUC) |
|---|---|---|
| Early-PD | De-novo | 0.92 |
| De-novo | Early-PD | 0.77 |
| Combined | | 0.81 |

**Table 4. Cross validation performance for nQi in the Early-PD and De-novo datasets.** The nQi scores for the combined dataset are generated by combining the output of the cross-validation of Early-PD and De-novo dataset, without any further training. The AUC scores allow a reliable comparison of the performance even when the number of PD and Controls is not fully balanced as in the Early-PD and De-novo datasets evaluated independently.

Computer use in the various age groups is an important factor to take into consideration for the applicability of our approach, a factor that is especially important given that older subjects are generally less likely to be computer users. Nonetheless, the United States Census Bureau estimates that the percentage of individuals owning a computer has grown to 71% in the 65+ age group as of 2013[29]. As this figure grows, the adoption of an approach such as the one described in this paper increases in feasibility.

The use of the natural interaction with commodity digital devices as a data source brings significant advantages. Data can be captured at home with a frequency much higher of the current standard of care. Furthermore the data capturing platform can be deployed easily at large scale and at a low cost. Additionally, high frequency at-home data collection addresses the problem of the artificial circumstances created during a consultation with a physician: it is not uncommon for patients to have unrepresentative scores in motor tests, either because of the Hawthorne effect[30] or because of the alteration of the timing of their medication to ensure that they arrive in good shape to the visit[31]. A data collection system that integrates with the normal use of a keyboard will also enable high compliance for all subjects who use digital devices for their job or as a pastime. The work described here provides a first step towards enabling that future, by showing in the limited context of a typing test done in a clinic for a small cohort, a good ability to discriminate between PD and control, and by doing so for the challenging case of an early PD population.

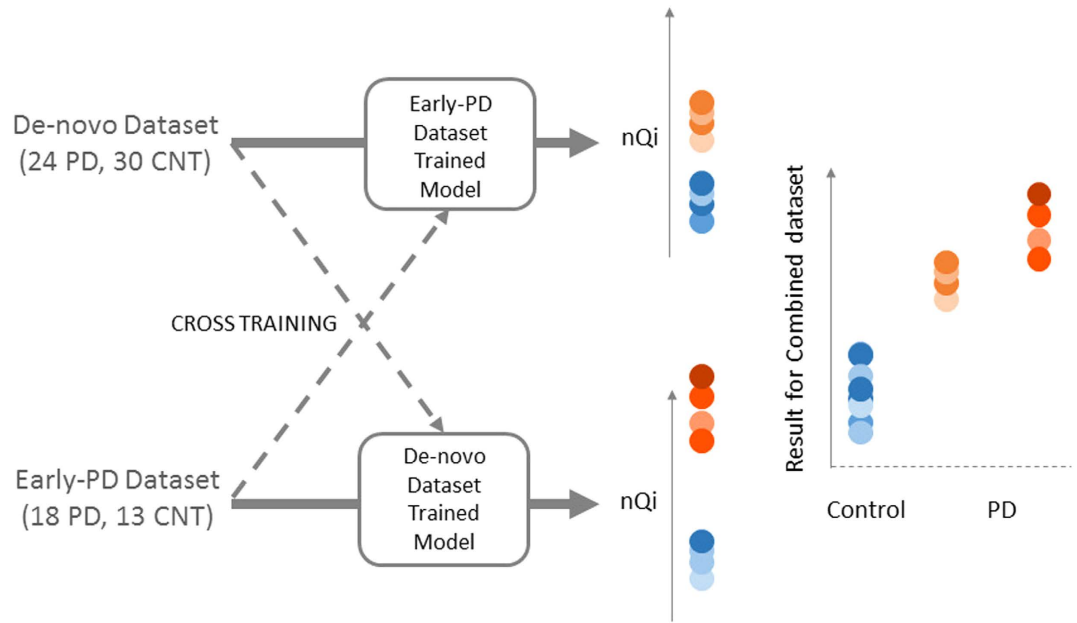

**Figure 4. Cross validation strategy.** The nQi scores for the Early-PD dataset are generated by training our ensemble regression model (see Methods) on the De-novo dataset, while the nQi scores for the De-novo dataset are generated by training on the Early-PD dataset. The nQi scores for the combined dataset are generated by combining the previous outputs without any further training.

**Experimental Procedures.** All the experimental protocols were approved by the Massachusetts Institute of Technology, USA (Committee on the Use of Humans as Experimental Subjects approval no. 1402006203 and no. 1412006804), Hospital 12 de Octubre, Spain (no. CEIC:14/090) and Hospital Clinico San Carlos, Spain (no. 14/136-E) and HM Hospitales, Spain (no. 15.05.796-GHM). Informed consent was obtained from all subjects involved in the study. All the experiments and recruitment were carried out in accordance with the relevant institutional guidelines.

**Early-PD Dataset.** This dataset consisted of 31 subjects, 18 early PD cases, i.e. patients without axial signs (Hoehn-Yahr stages I and II), without motor fluctuations and with a confirmed diagnosis for less than 5-years and 13 healthy spouses without any sign of parkinsonism as controls[32] (one PD subject 6 years since diagnosis was included as early-PD because of very mild PD motor signs). Only participants who self-reported that they used a laptop or desktop computer for at least thirty minutes per day and completed all visits were considered eligible. We excluded participants with cognitive impairment or dementia, subjects with upper limb functional limitation, antipsychotics/sedative users and participants with sleep disorders. The subjects were recruited from two movement disorder units in Madrid (Spain). Table 3 shows demographic and clinical information of this cohort.

Each subject was asked to visit a movement disorder unit twice where the motor tests and clinical evaluation were performed, with 7 to 30 days between each visit. Patients taking levodopa, a symptomatic relief medication for PD, were asked to refrain from taking the medication for 18 hours before the visit.

**De-novo Dataset.** This cohort was recruited as part of an on-going longitudinal PD study. It consisted of 54 subjects, 24 de-novo PD cases, i.e. newly diagnosed, drug naïve patients, and 30 healthy controls without any sign of parkinsonism. The PD subjects were recruited from 8 different health institutions in Madrid (Spain), the controls were mainly patient's spouses or subjects connected to the PD community. Only participants who self-reported that they used a laptop or desktop computer for at least thirty minutes per day were considered eligible. We excluded participants with cognitive impairment or dementia, subjects with upper limb functional limitation, antipsychotics/sedative users and participants with sleep disorders. Table 3 shows demographic and clinical information of this cohort.

**Tests Performed.** Each visit involved a clinical evaluation, finger tapping tests and our typing test. The clinical evaluation was undertaken by a movement disorder specialists who also filled-in the motor section in the Unified Parkinson's Rating Scale (UPDRS-III)[11]. In the Early-PD dataset it was performed by two blinded specialists while in the De-novo dataset by a single one.

The finger tapping tests are a common way to quantify upper limbs dexterity in clinical studies. In the "single key tapping" test, subjects repeatedly pressed a single button for 60 seconds, as fast as possible, first with their dominant hand, then with the non-dominant hand. The final score was the average number of buttons pressed between the two hands. In the "alternating finger tapping" test, the subjects had to alternatively press two buttons, with a distance of approximately 25 cm between the two, with their index finger. The test was repeated for both hands and the final score was the average number of buttons pressed between the two hands. The "alternating finger tapping" test was introduced while the early-PD study was ongoing, because of this 5 PD subjects and 4 controls could only be measured with the "single key tapping" test.

In the typing test, the subjects transcribed a folk tale on standard word processor. The folk tale was randomly selected from a collection avoiding repetitions of the same text for each given subject. This was to limit the learning effect due to the content of the text. The subjects were instructed to type as they normally would do at home and they were left free to correct typing mistakes only if they wanted to. All subjects typed for an average of 14 minutes (2.9 std) with a standard word processor on a Lenovo G50-70 i3-4005U with 4MB of memory and a 15 inches screen running Manjaro Linux. In the background a custom piece of software recorded the timestamps of each key press and depress, stored it in memory and sent it to a remote database at the end of the writing task.

The timing resolution of the key acquisition software was evaluated by injecting a series of software generated key presses and releases into the operating system event queue. A stream of two consecutive events (key-press, key-release) was generated every 100 milliseconds for a total running time of 15 minutes. We measured a temporal resolution of 3/0.28 (mean/std) milliseconds. The versions for Windows and Macintosh can be downloaded at https://www.neuroqwerty.com.

PD patients and controls of the Early-PD dataset were tested twice, while the subjects in the De-novo dataset once. In order to make our experiments consistent, we averaged the clinical and motor tests scores across the two repetitions in the Early-PD dataset.

## Methods

We present a new computational algorithm able to generate a Parkinson's Disease motor index (nQi) that is used to classify subjects as having or not having (early) Parkinson's Disease from data obtained during a natural typing task. More specifically, the data source is a series of hold times, the time between pressing and releasing a key on a laptop keyboard. First, we introduce a new type of typing signal representation extending our previous work[23] with variance analysis features; then, we use an ensemble approach based on linear $\varepsilon$-Support Vector Regression to generate nQi scores. Figure 1 shows a visual representation of the algorithmic pipeline.

**Signal Representation.** Let the vector $a[t]$ represent continuous-time stochastic process of key hold times where $t$ is the time at which each key has been pressed. We consider only the keys for which we expect a short hold time, i.e. alphanumeric characters, symbols and space bar. We define a square window $\omega$, such that:

$$\omega[n] = \begin{cases} 1, & \text{if } 0 \leq n < N^w \\ 0, & \text{otherwise} \end{cases} \quad (1)$$

where $N^w$ is the size of the window expressed in seconds. In our experiments we used $N^w = 90$. Then, it is possible to partition the hold time signal $a$ with non-overlapping square windows as follow: $B_i = a[t]\omega[t - iN^w]$ where $t$ is time, $B_i$ is a vector containing the ordered list of HT samples and $i$ is a positive integral number which serves as index to the list of vectors. In order to account for the sparsity of the hold times signal, i.e. do not type continuously but in unpredictable bursts, all $B_i$ that have less than $N^w/3$ key presses are removed from the set. Let us define a feature vector for each $B_i$:

$$x_i = [\nu^{\text{out}}, \nu^{\text{iqr}}, \nu^{\text{de}}, \nu^{\text{hst}_0}, \nu^{\text{hst}_1}, \nu^{\text{hst}_2}, \nu^{\text{hst}_3}]^T \quad (2)$$

where $\nu^{\text{out}}$ is the number of outliers in $B_i$ divided by the number of elements in $B_i$. An outlier is defined as a HT more than 1.5 interquartile ranges below the first quartile or above the third quartile; $\nu^{\text{iqr}}$ is a measure of the $B_i$ distribution skewness described as $(q_2 - q_1)/(q_3 - q_1)$, and $q_n$ is the nth quartile; $\nu^{\text{hst}_n}$ represents the nth bin of the $B_i$ equally-spaced normalized histogram, i.e. an approximation of the probability density function, with 4 bins from 0 to 0.5 seconds; $\nu^{\text{de}}$ is a metric of finger coordination during two consecutive keystrokes. It is measured as $d_1 - p_2$, where $d_1$ is the depress event of the first key and $p_2$ is the press event of the second key. If $(d_1 - p_2) < 0$, then $\nu^{\text{de}} = 0$.

**Ensemble regression.** We designed an ensemble learning approach composed of a set of base models $F: \{f_m | m \geq 0 \land m < N^m\}$ where $N^m$ is the total number of models, which in our experiments is 200. Each model $f_m$ receives as input an independent feature vector $x_i$ and performs a linear regression step with $\varepsilon$-Support Vector Regression as follow:

$$\begin{aligned} y'_m &= f_m(x_{i'}) \\ &= b_m + w_m^T x_{i'} \end{aligned} \quad (3)$$

$f_m(x_{i'})$ is a linear $\varepsilon$-Support Vector Regression model implemented in LibSVM[33]; The result $y'_m$ is a partial estimation of the nQi score. The nQi for each $x_i$ is calculated by applying all the regression models in $F$ on the $x_i$ vector and then calculating the median score. Using a Bagging strategy, we generated a different set of $w_m$ and $b_m$ coefficients for each $f_m$ during the training phase. Bagging allows the creation of $N^m$ views of the training dataset by generating multiple sets (or bootstrap samples) via random sampling with replacement. This approach reduces the variance in the nQi score and further limits chances of overfitting[34].

All regression models $f_m$ were trained keeping the Support Vector parameters $C$ and $\varepsilon$ fixed. Formally, each $f_m$ is trained by minimizing the standard convex minimization problem for the set of $l$ training vector and target values $\{(x_{i'=1}, (z_{i'=1})), \ldots, (x_{i'=l}, (z_{i'=l}))\}$ in the bootstrap sample:

$$\begin{aligned}
&\underset{w,b,\xi,\xi^*}{\text{minimize}} && \frac{1}{2}w^T w + C\sum_{i'=1}^{l}(\xi_{i'}, \xi_{i'}^*) \\
&\text{subject to} && w^T x_{i'} + b - z_{i'} \leq \varepsilon + \xi_{i'}, \\
&&& z_{i'} - w^T x_{i'} - b \leq \varepsilon + \xi_{i'}^*, \\
&&& \xi_{i'}, \xi_{i'}^* \geq 0.
\end{aligned} \quad (4)$$

where $C > 0$ and $\varepsilon > 0$; $z_{i'}$ are the target values, i.e. the normalized UPDRS scores; $l$ is the size of each bootstrap sample; $\xi$ is the $\varepsilon$-insensitive loss function[34]. The $m$ identifying each separate model in $F$ has been omitted from the variables in the minimization problem for readability.

In order to estimate the regressor parameters $C$ and $\varepsilon$, we leveraged an external dataset composed of typing signals of 7 PD subjects and 18 controls not present in the De-novo or Early-PD datasets. This parameter estimation dataset was created by joining the data from our previous work[23] and the typing data that was collected but could not be used for the De-novo or Early-PD dataset because of the exclusion criteria. The parameters were estimated with a grid search approach that maximized the AUC estimated with a leave-one-out cross validation strategy for each set of parameters. The parameters found were $C = 0.094$ and $\varepsilon = 0.052$.

Unless explicitly mentioned, all statistical analyses were performed with a single score per subject. This was calculated by averaging together all the independent nQi scores computed during the typing task as shown in Fig. 1(3–4).

## References

1. Hirtz, D. *et al.* How common are the "common" neurologic disorders? *Neurology* **68,** 326–337 (2007).
2. de Lau, L. M. L., Koudstaal, P. J., Hofman, A. & Breteler, M. M. B. Subjective Complaints Precede Parkinson Disease. *Arch. Neurol.* **63,** 362–365 (2006).
3. Ross, G. W., Abbott, R. D., Petrovitch, H., Tanner, C. M. & White, L. R. Pre-motor features of parkinson's disease: the honolulu-asia aging study experience. *Parkinsonism Rel. Disord.* **18,** 199–202 (2012).
4. Berg, D. *et al.* The PRIPS study: Screening battery for subjects at risk for Parkinson's disease. *Eur. J. Neurol.* **20,** 102–108 (2013).
5. Macleod, A. D., Taylor, K. S. M. & Counsell, C. E. Mortality in Parkinson's disease: a systematic review and meta-analysis. *Mov. Disord.* **29,** 1615–22 (2014).
6. Forsaa, E. B., Larsen, J. P., Wentzel-Larsen, T. & Alves, G. What predicts mortality in Parkinson disease?: a prospective population-based long-term study. *Neurology* **75,** 1270–6 (2010).
7. Löhle, M., Ramberg, C.-J., Reichmann, H. & Schapira, A. H. V. Early versus delayed initiation of pharmacotherapy in Parkinson's disease. *Drugs* **74,** 645–57 (2014).
8. Willis, a. W., Schootman, M., Evanoff, B. a., Perlmutter, J. S. & Racette, B. a. Neurologist care in Parkinson disease: a utilization, outcomes, and survival study. *Neurology* **77,** 851–7 (2011).
9. Lang, A. E. Clinical trials of disease-modifying therapies for neurodegenerative diseases: the challenges and the future. *Nat. Med.* **16,** 1223–1226 (2010).
10. Streffer, J. R. *et al.* Prerequisites to launch neuroprotective trials in Parkinson's disease: an industry perspective. *Mov. Disord.* **27,** 651–5 (2012).
11. Martínez-Martín, P. *et al.* Unified Parkinson's Disease Rating Scale characteristics and structure. The Cooperative Multicentric Group. *Mov. Disord.* **9,** 76–83 (1994).
12. Little, M., Wicks, P., Vaughan, T. & Pentland, A. Quantifying short-term dynamics of Parkinson's disease using self-reported symptom data from an Internet social network. *J. Med. Internet Res.* **15,** e20 (2013).
13. von Campenhausen, S. *et al.* Costs of illness and care in Parkinson's Disease: An evaluation in six countries. *Eur. Neuropsychopharmacol.* **21,** 180–191 (2011).
14. Stamford, J., Schmidt, P. & Friedl, K. What Engineering Technology Could Do for Quality of Life in Parkinson's Disease: a Review of Current Needs and Opportunities. *IEEE J. Biomed. Health Inform.* 1–11 (2015).
15. Sánchez-Ferro, A. *et al.* New methods for the assessment of Parkinson's disease (2005 to 2015): A systematic review. *Mov. Disord.* (2016) (in press).
16. Taylor Tavares, A. L. *et al.* Quantitative measurements of alternating finger tapping in Parkinson's disease correlate with UPDRS motor disability and reveal the improvement in fine motor control from medication and deep brain stimulation. *Mov. Disord.* **20,** 1286–98 (2005).
17. Maetzler, W., Domingos, J., Srulijes, K., Ferreira, J. J. & Bloem, B. R. Quantitative wearable sensors for objective assessment of Parkinson's disease. *Mov. Disord.* **28,** 1628–1637 (2013).
18. Horak, F. B. & Mancini, M. Objective biomarkers of balance and gait for Parkinson's disease using body-worn sensors. *Mov. Disord.* **28,** 1544–1551 (2013).
19. Arora, S. *et al.* Detecting and monitoring the symptoms of Parkinson's disease using smartphones: a pilot study. *Parkinsonism Rel. Disord.* **21,** 2015–2018 (2015).
20. Ahmad, N., Szymkowiak, A. & Campbell, P. a. Keystroke dynamics in the pre-touchscreen era. *Front. Hum. Neurosci.* **7,** 835 (2013).
21. Banerjee, S. & Woodard, D. Biometric authentication and identification using keystroke dynamics: A survey. *J. Pattern Recognition Res.* **7,** 116–139 (2012).
22. Austin, D., Jimison, H., Hayes, T., Mattek, N. & Pavel, M. Measuring motor speed through typing: a surrogate for the finger tapping test. *Behav. Res. Methods* **43,** 903–909 (2011).

23. Giancardo, L., Sánchez-Ferro, A., Butterworth, I., Mendoza, C. S. & Hooker, J. M. Psychomotor impairment detection via finger interactions with a computer keyboard during natural typing. *Sci. Rep.* **5,** 9678 (2015).
24. Rempel, D. & Dennerlein, J. A method of measuring fingertip loading during keyboard use. *J. Biomech.* **27,** 1101–1104 (1994).
25. Kuo, P.-L., Lee, D. L., Jindrich, D. L. & Dennerlein, J. T. Finger joint coordination during tapping. *J. Biomech.* **39,** 2934–42 (2006).
26. Shimoyama, I., Ninchoji, T. & Uemura, K. The finger-tapping test. A quantitative analysis. *Arch. Neurol.* **47,** 681–684 (1990).
27. Jobbágy, Á., Harcos, P., Karoly, R. & Fazekas, G. Analysis of finger-tapping movement. *J. Neurosci. Methods* **141,** 29–39 (2005).
28. O'Boyle, D. J., Freeman, J. S. & Cody, F. W. J. The accuracy and precision of timing of self-paced, repetitive movements in subjects with Parkinson's disease. *Brain* **119,** 51–70 (1996).
29. File, T. & Ryan, C. Computer and Internet use in the United States: 2013. *Current Population Survey Reports. US Census Bureau* (2014).
30. McCarney, R. *et al.* The Hawthorne Effect: a randomised, controlled trial. *BMC Med. Res. Methodol.* **7,** 1–8 (2007).
31. Cramer, J. A., Scheyer, R. D. & Mattson, R. H. Compliance declines between clinic visits. *Arch. Intern. Med.* **150,** 1509–1510 (1990).
32. The Parkinson Study Group. Effects of tocopherol and deprenyl on the progression of disability in early Parkinson's disease. *N. Engl. J. Med.* **328,** 176–83 (1993).
33. Chang, C.-C. & Lin, C.-J. LIBSVM: A Library for Support Vector Machines. *ACM Trans. Intell. Syst. Technol.* **2,** 27 (2011).
34. Bishop, C. *Pattern Recognition and Machine Learning* (Springer, 2007) 2nd edn.
35. Schisterman, E. F., Perkins, N. J., Liu, A. & Bondell, H. Optimal cut-point and its corresponding Youden Index to discriminate individuals using pooled blood samples. *Epidemiology* **16,** 73–81 (2005).

### Acknowledgements

This research is being financially supported by the Comunidad de Madrid, Fundacion Ramon Areces and The Michael J Fox Foundation for Parkinson's research (grant number 10860). We thank the M + Vision faculty for their guidance in developing this project. We also thank our many clinical collaborators at MGH in Boston, at "12 de Octubre", Hospital Clinico and Centro Integral en Neurociencias HM CINAC in Madrid for their insightful contributions.

### Author Contributions

L.G., A.S.-F., I.B., C.S.M., M.L.G. and J.A.O. conceived the experiments; L.G. and T.A. developed and tested the algorithms; L.G. and R.J.E. analyzed the results; L.G., M.L.G., R.J.E. and A.S.F. interpreted the findings; A.S.-F., P.M. and M.M. conducted the human study. All authors discussed the results and contributed to the manuscript.

### Additional Information

**Supplementary information** accompanies this paper at http://www.nature.com/srep

**Competing financial interests:** The authors declare no competing financial interests.

**How to cite this article**: Giancardo, L. *et al.* Computer keyboard interaction as an indicator of early Parkinson's disease. *Sci. Rep.* **6,** 34468; doi: 10.1038/srep34468 (2016).